\documentclass[journal]{IEEEtran}
\usepackage[colorlinks,urlcolor=blue,linkcolor=blue,citecolor=blue]{hyperref}
\usepackage{cite}
\usepackage{amsmath,amssymb,amsfonts}
\usepackage{algorithmic}
\usepackage{graphicx}
\usepackage{textcomp}
\usepackage{color,array}
\usepackage{multirow}
\usepackage{multicol}
\usepackage{graphicx}
\usepackage{caption}
\usepackage{footnote}
\usepackage{float}
\usepackage{tabularx}
\usepackage{booktabs}
\usepackage[caption=false]{subfig}
\usepackage{array}
\newcolumntype{x}[1]{>{\centering\arraybackslash}m{#1}}
\usepackage{scalerel}
\usepackage{tikz}

\usepackage{hyperref} %<--- Load after everything else
\definecolor{CRED}{RGB}{255,0,0}
\definecolor{CGREEN}{RGB}{0,255,0}
\definecolor{CBLUE}{RGB}{0,0,255}

\begin{document}

\title{Dense Regression Activation Maps For Lesion Segmentation in CT scans of COVID-19 patients} 

\author{Weiyi Xie, \and
        Colin Jacobs, \and
        Jean-Paul~Charbonnier, \and
        Bram van Ginneken% <-this % stops a space
\thanks{(\textit{Corresponding author: Weiyi Xie, e-mail: weiyi.xie@radboudumc.nl})
}%
% \thanks{This paper was submitted on 18th November, 2021 for review. This work was supported by the Dutch Lung Foundation under the project 5.1.17.171.}%
\thanks{}%
\thanks{Weiyi Xie and Colin Jacobs and Bram van Ginneken are with the Diagnostic Image Analysis Group, Department of Radiology and Nuclear Medicine, Radboudumc, 6525 GA Nijmegen, The Netherlands (weiyi.xie@radboudumc.nl; colin.jacobs@radboudumc.nl; bram.vanginneken@radboudumc.nl).}
\thanks{Jean-Paul Charbonnier is with Thirona, 6525 EC Nijmegen, The Netherlands (jeanpaulcharbonnier@thirona.eu).}} %Format necessary to be eligible for RIHS Science Award

\markboth{}%IEEE TRANSACTIONS ON MEDICAL IMAGING}%
{W.Xie \MakeLowercase{\textit{et al.}}: Dense Regression Activation Maps For Lesion Segmentation in CT scans of COVID-19 patients}

\maketitle

\begin{abstract}
Automatic lesion segmentation on thoracic CT enables rapid quantitative analysis of lung involvement in COVID-19 infections. However, obtaining a large amount of voxel-level annotations for training segmentation networks is prohibitively expensive. Therefore, we propose a weakly-supervised segmentation method based on dense regression activation maps (dRAMs). Most weakly-supervised segmentation approaches exploit class activation maps (CAMs) to localize objects. However, because CAMs were trained for classification, they do not align precisely with the object segmentations. Instead, we produce high-resolution activation maps using dense features from a segmentation network that was trained to estimate a per-lobe lesion percentage. In this way, the network can exploit knowledge regarding the required lesion volume. In addition, we propose an attention neural network module to refine dRAMs, optimized together with the main regression task. We evaluated our algorithm on 90 subjects. Results show our method achieved 70.2\% Dice coefficient, substantially outperforming the CAM-based baseline at 48.6\%. 
\end{abstract}


\section{Introduction}
\IEEEPARstart{T}{HE} coronavirus disease 2019 (COVID-19) has been declared a global pandemic since March, 2020. Infected cases have reached over 252 million worldwide, with more than five million deaths. Unfortunately, both numbers are still increasing. To reduce the fatality rate, effective diagnosis and treatment planning are essential. As COVID-19 mainly damages the lungs of infected subjects, Computed Tomography (CT) of the chest plays a critical role in rapid diagnosis and progression monitoring of COVID-19 infection. Based on chest CT analysis, standardized CT scoring systems, such as the COVID-19 Reporting and Data System (CO-RADS) \cite{Prok20}, were defined to quantify the degree of suspicion of COVID-19 according to CT findings into 1-5 scores with an increasing level of suspicion.

Similarly, a CT severity scoring system \cite{Li20} was designed to assess the extent of parenchymal involvement of the disease. These scoring systems may be applied more accurately, objectively, and rapidly when automatic segmentation of infected areas (lesions) is available. Therefore, this work aims at developing an algorithm that can automatically segment lesions related to COVID-19 on chest CT scans. 

One of the major obstacles of semantic segmentation is the difficulty of acquiring a large amount of voxel-wise annotations for training as manual outlining in high-resolution 3D scans is extremely laborious.  Therefore, we present a novel weakly-supervised segmentation method that only requires lobe-wise severity scores, such as routinely reported using the CT severity scoring system, to supervise training. Using only these lobe scores, we aim to produce high-resolution lesion segmentation maps.

Weakly-supervised segmentation (WSS) has been extensively studied in recent years, where reference standards can be provided using scribbles \cite{Ji19}, or surface points \cite{Roth19}. Both these approaches seek a trade-off between annotation efforts and location information in need. However, because COVID-19 CT abnormalities often have bilateral lung involvement with a peripheral and diffuse distribution \cite{Shi20a}, annotating scribbles or extreme points could still be demanding. A less demanding approach is to label the entire image volume or only regions within a volume. Early WSS methods using image-level labels were based on multi-instance learning frameworks~\cite{Pinh15} and the expectation-maximization algorithm~\cite{Papa15}. Compared to these early works, approaches based on class activation maps (CAMs) \cite{Huan18a,Ahn18,Wang20,Wei17} significantly improved the segmentation performance on major benchmarks \cite{Wang20}. CAMs were originally generated at low resolution, usually from the features before global pooling or fully connected layers in training a classification network. Low-resolution CAMs do not provide local details. 

To obtain high-resolution activation maps, MS-CAM \cite{Ma20} exploited both low- and high-resolution convolution features in a multi-scale framework. High-resolution CAMs may produce segmentation maps with local details. However, they still do not necessarily align with object segmentation, because naturally CAMs only reflect discriminative regions responsible for classification. This may causes certain objects or parts being ignored in the CAM, as long as the classification was made correctly. To mitigate this issue, research efforts were made to expand or refine CAMs, often using the original CAMs as the initial object cues or seed regions. Wei et al.~\cite{Wei17} proposed to progressively erase already-found object maps and force the network to discover new and complement CAM regions at later runs. A similar iterative approach was applied on lesion localisation in color fundus images \cite{Gonz20}. A seeded region growing module was proposed in \cite{Huan18a} to expand CAMs to cover the complete object boundaries. AffinityNet \cite{Ahn18} exploited inter-pixel affinities as the transition probability matrix and applied random walks to expand and refine CAMs.\par

Instead of relying on CAMs, we obtain object cues based on dense regression activation maps (dRAMs) by training a segmentation network for regressing the per-lobe lesion percentage. dRAMs can provide high-resolution segmentation maps. Moreover, regression training allows the network to be aware of the object size, which was not done in previous work that relied on training with categorical labels. As described in \cite{Less20}, lobe-wise severity scores provided by radiologists can be translated into intervals of lesion percentage per lobe. We propose a novel interval regression loss to enforce the predicted lesion percentage to fall in a particular range to use these intervals as reference standards.

Furthermore, we introduce an attention module for revising dRAMs, trained together with the regression task. This refinement module updates each location in dRAMs using information from its local neighbors, according to voxel-wise affinities embedded in convolution features. This process mimics the random-walker in \cite{Ahn18} and seeded-region growth in \cite{Huan18a}, but contrary to those approaches, it was trained end-to-end in our framework. 

Our key contributions are as follows: 1) we propose a lesion segmentation framework that produces high-resolution segmentation maps using only lobe-wise labels for training; 2) we convert the lesion segmentation problem to regression of the per-lobe lesion percentage. The regression problem is solved using a proposed interval regression loss. These ideas are generic and can be extended to other WSS problems; 3) we refine the dense regression activation maps using a proposed attention neural network module trained together with the main regression target and show this obviates the need for ad hoc post-processing steps. 

\subsection{Related works}
Several recent works on COVID-19 lesion segmentation attempted to reduce the demand for voxel-level supervision in training. Fan et al.~\cite{Fan20} proposed a semi-supervised training strategy that requires a few labeled images to train the initial segmentation model and leverages primarily unlabeled data to fine-tune the model progressively. Laradji et al.~\cite{Lara20} proposed to use point-level labels in an active learning schema to generate lesion segmentation maps. Yao et al.~\cite{Yao20} superimposed synthesized lesions on healthy CT scans to train networks to separate lesions from other structures. Xu et al.~\cite{Xu20a} relied on scan-level labels and fractional voxel-level labels to train a generative adversarial learning framework for segmenting COVID-19 lesions. Wang et al.~\cite{Wang20a} proposed to train a binary classifier based on the presence of COVID-19 on CT scans and used the classifier to generate CAMs for detecting lesions.  

Our approach is closely related to CAMs-based WSS approaches in three building blocks: 1) the generation of CAMs by training a convolution neural network, often for a classification task. 2) regularization, to stabilize training. 3) CAM refinement, which is often needed because CAMs do not necessarily align with object boundaries. \par 
In these three aspects, our approach generates dense class regression maps (dRAMs) by training a segmentation network towards a regression target. Meanwhile, we introduce self-supervised learning by equivariant regularization \cite{Wang20} for improving the consistency of extracted dRAMs over various affine transformed input images. In terms of the refinement, our method is motivated by \cite{Ahn18}, which revised CAMs by a random-walker in post-processing steps where the transitional probabilities were derived from learned local affinities. A similar but end-to-end solution can be found in \cite{Wang20} where an attention-based neural network module was proposed to capture global affinities for refining CAMs. In our method, we capture affinities via an attention module inspired by \cite{Wang20} but computed within a local neighborhood, similar to that in \cite{Ahn18}.

\section{Data}

\begin{table}
 \caption{
  The distribution of CO-RADS scores and lobe severity scores across the training and test sets. CO-RADS score 1-6 indicates the level of suspicion for COVID-19 positive disease, ranging from very low, low, equivocal, high, very high, and confirmed PCR positive, respectively. Lobe severity scores indicate the extent of lobe-wise involvement of COVID-19 infection.}
 \centering
\caption*{(a) CO-RADS scores}
 \begin{tabular}{lll}
  \toprule
  CO-RADS     & \#subjects for training 
  & \#subjects for testing \\
  \midrule
  1     & 34 & 0     \\
  2     & 35 & 0     \\
  3     & 96 & 19     \\
  4     & 47 & 21     \\
  5     & 68 & 35     \\
  6    & 20 & 15     \\
  \bottomrule
Total & 300 & 90\\
  \bottomrule
\end{tabular}
\vspace*{0.1 cm}
\caption*{(b) Lobe severity scores}
\begin{tabular}{p{3cm}p{1.5cm}p{1.5cm}}
  \toprule
  severity scores \newline(percentage per lobe) & \#training lobes
  & \#testing lobes \\
  \midrule
  0 (0\%)     & 535 &  64     \\
  1 (1-5\%)    & 327 &  90     \\
  2 (5-25\%)    & 308 &  154     \\
  3 (25-50\%)    & 185 &  86     \\
  4 (50-75\%)    & 110 &  42     \\
  5 (75-100\%)   & 35 &  14     \\
  \bottomrule
Total Lobes & 1500 & 450 \\
\bottomrule
\end{tabular}
\label{tab:data}
\end{table}

We selected CT scans from patients who presented at the emergency wards of the Radboud University Medical Center, the Netherlands, from March to September 2020. Patients were referred for CT imaging because of suspicion of moderate to severe COVID-19 pneumonia. The ethical review board approved the retrospective and anonymous collection of this data (Radboudumc CMO2016-3045, Project 20027). All CT scans were obtained with a low-dose thin slice protocol without administration of contrast (details in \cite{Less20}). 

Following the guidelines of the Dutch Radiological Society \cite{Prok20}, the radiology report for each scan contained CO-RADS and lobe-wise severity scores. CO-RADS 1 is defined as a scan that is normal or has non-infectious etiologies, and thus a very low level of suspicion for COVID-19. CO-RADS 2 indicates the CT-scan has features typical for infections other than COVID-19. CO-RADS 3 indicates equivocal findings: features compatible with COVID-19 but also with other diseases. CO-RADS 4 and 5 indicate a high and very high level of COVID-19 suspicion, respectively. CO-RADS 6 was given to scans from patients already known to be positive for COVID-19 with reverse transcription-polymerase chain reaction (RT-PCR) tests at the time of reporting. Lobe-wise severity scores indicate the extent of lobar involvement of the COVID-19 infection. A score from 0 to 5 is assigned to each lobe according to the visually assessed lesion percentage of that lobe. The mapping between lobe-wise severity score and lesion percentage per lobe can be found in Table \ref{tab:data}(b). We used lobe severity scores to generate the weak labels in training our algorithm.

\subsubsection{Data Selection and Partitioning} \label{data_selection}
We randomly selected 390 subjects (shuffled into 300 for training and 90 for testing). This selection included all subjects that were available when this project started. A single scan was used for each subject. Thirty subjects in the training set were used as the validation set during model development to prevent over-fitting. The distribution of CO-RADS and lobe severity scores are provided in Table \ref{tab:data} (a and b). 

\subsubsection{Reference Standard}\label{label_protocol}
For evaluating our method, lesion segmentation references on 90 test scans were obtained from Thirona (Nijmegen, the Netherlands), a medical image analysis service company specializing in chest CT analysis. First, lung parenchyma regions with a higher attenuation were identified by thresholding and morphologic operations. Next, automatic methods were used to suppress vessels and airways. Following the radiology report, a certified image analyst with at least one year of experience corrected the segmentations obtained by the automatic methods. The analysts also labeled segmented lesions into the ground glass, consolidation, and mixed type, to allow us to evaluate segmentation performance for different lesion subtypes. The analyst could consult a radiologist in cases of doubt during the annotation process.

\section{Methods}
\begin{figure*}[!t]
	\centering
	\includegraphics[width=0.9\linewidth]{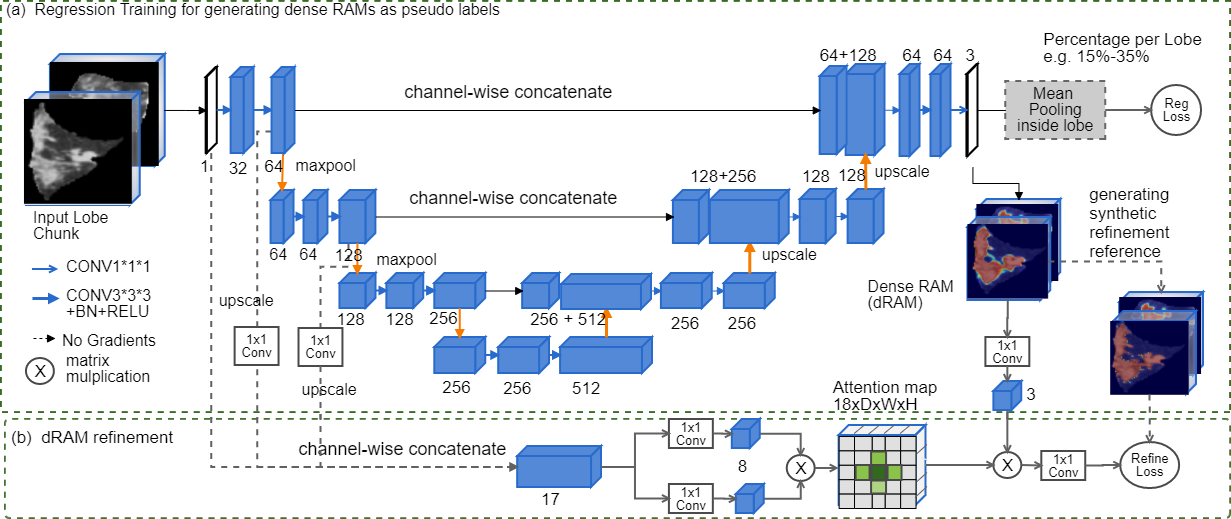}
	\caption{Overview of the proposed weakly-supervised segmentation framework. (a) shows the backbone 3D U-net \cite{Cice16} is not trained with segmentation masks, only performing a regression task, i.e., predicting the percentage of lesions per lobe. The final dense features were used to generate a dense regression activation map (dRAM), corresponding to the lesion segmentation. (b) shows the attention module for dRAM refinement, using synthetic segmentation reference generated with lesion candidates and dRAMs. }
	\label{fig:overview}
\end{figure*}

\subsection{Weakly-supervised segmentation framework}
The overview of the proposed weakly-supervised lesion segmentation framework is shown in Fig.~\ref{fig:overview}. We train a regression network to predict the lesion percentage per lobe, and in the process, we generate the dense regression activation maps (dRAMs). The network is trained using an interval regression loss because the exact lesion percentage per lobe is not available for training. Instead, we only know the range of lesion percentage from lobe-wise severity scores, as given in Table \ref{tab:data} (b). For example, radiologists would label a lobe with a severity score of 1 when the estimated lesion percentage in that lobe is between 1\% and 5\%. All lobe-wise severity scores were translated into lesion percentage intervals. Before using these intervals as the regression target, we calibrate the lesion percentage intervals using a candidate proposal technique to reduce the annotation errors.\par 
Moreover, equivariant regularization is used to improve the consistency of dRAMs among various affine-transformed input images. Finally, dRAMs are refined in an auxiliary training task using a proposed attention mechanism in addition to the regression and regularization. The following subsections elaborate on each of these steps.

\subsubsection{Low-level features for lesion candidate proposal}\label{sec:regio_pro}
Low-level features were commonly used in weakly-supervised semantic segmentation. For example, Wei et al.~\cite{Wei17a} used low-level features-based saliency maps as the reference for training the initial segmentation network. We propose to isolate high attenuation areas in the lung by applying Otsu's threshold \cite{Otsu79} separately for each lobe. Meanwhile, we coarsely suppress vessels via vessel enhancement filtering \cite{Fran98}. The high attenuation lung regions after vessel suppression are marked as candidate lesions. 

\subsubsection{Regression training for generating dRAMs}\label{sec:reg_loss}
The dense regression activation map is generated by training a regression network for predicting the lesion percentage per lobe. We adopt the 3D U-Net \cite{Cice16} as the regression network (shown in Fig.~\ref{fig:overview} (a)). The 3D U-Net has three down-sampling layers in the encoding path, and each layer consists of two convolutions and a max-pooling operation. Following the down-sampling path, two more convolutions are used to double the number of convolution filters. In the up-sampling path, three layers are used to reconstruct the resolution, and each contains one tri-linear interpolation, followed by two convolutions to reduce the interpolation artifacts. Convolution kernels have $3\times 3 \times 3$ kernel size, a stride of 1 voxel, and zero padding. At the final upsampling layer, feature maps are dense (having the same resolution as the network input). Then a regression head with a single $1\times 1 \times 1$ convolution and a softmax layer is used to output a $C$-channel dense prediction map. The regression network takes a lobe chunk image as the input, which is cropped around each segmented lobe and resized to have a fixed spatial size $D \times W \times H$, where $D$, $W$, and $H$ are the number of slices, width, and height in the input lobe chunk image. The pulmonary lobe segmentation was performed using a publicly available algorithm \cite{Xie20}. For each lobe chunk input, areas outside the lobe of interest are masked out. Denoting the lobe chunk input image as $I$ and the regression network as $F(\cdot)$, forwarding $I$ through $F(\cdot)$ produces a $C \times D \times W \times H$ dimensional dense prediction map $F(I)$, referred to as the dense regression activation map (dRAM). In this paper, $C$ channels are channels corresponding to the background, vessels, and abnormal regions in ascending order. For any channel, we can compute the per-lobe percentage of the object corresponding to the channel by simply average-pooling $F(I)$ over all voxels within the lobe (lobe-wise mean pooling), assuming dRAMs are segmentation maps. Denoting the lobe-wise mean pooling as $P(\cdot)$, the parameters of $F(\cdot)$ can be trained given a target percentage range $(r_{l}, r_{u})$ via the interval regression loss $L_{INT}$:
\begin{equation}\label{eq:interval_regression_loss}
\begin{array}{l}
   \mbox{min:~} max(0, (P(F(I)) - 0.5*(r_{l} + r_{u}))^2-K ), \\
                \qquad \quad K = (0.5 * (r_{l} – r_{u}))^2.
\end{array}
\end{equation}
Note that $P(F(I))$, $r_{l}$, and $r_{u}$ are float values between 0 and 1 representing a percentage. $L_{INT}$ can also be interpreted as the quadratic version of the piecewise linear loss function that minimizes $|P(F(I))-r_{l}| + |P(F(I))-r_{u}|-|r_{u}-r_{l}|$ to force the prediction $P(F(I))$ falls into $(r_{l}, r_{u})$. \par
The initial lesion percentage range ($r_{l}^{*}$,$r_{l}^{*}$) translated from lobe-wise severity scores (See Table \ref{tab:data} (b) for the definition) may contain measurement errors due to inaccurate estimation of lesion percentage when annotating severity scores. To reduce these measurement inaccuracies, we first calculate an estimation of per-lobe lesion percentage $p^{*}$ using candidate lesions (\ref{sec:regio_pro}) as the volume of the candidate lesions divided by the volume of the corresponding lobe. Then we use $p^{*}$ to calibrate the initial range via:
\begin{equation}\label{eq:calibration_interval}
\begin{array}{l}
    r_{l} = max(min(r_{l}^{*}, p^{*} - 0.05), 0.0) . \\
    r_{u} = min(r_{u}^{*}, p^{*} + 0.05)
\end{array}
\end{equation}
After the calibration, $r_{l}$ and $r_{l}$ are aware of the percentage of candidate lesions. The calibration can avoid overestimating the percentage of lesions during labeling lobe severity scores when both $r_{l}^{*}$, $r_{l}^{*}$ or one of them are beyond the percentage of candidate lesions.
We added 5\% tolerance for the possible errors in lesion candidate proposals ($p^{*} \pm 0.05$). 

\subsubsection{Equivariant regularization}\label{sec:er_loss}
Training on weak labels may lead to a trivial segmentation solution because no location information is given. Therefore, regularization techniques are commonly used to stabilize training. Wang et al.~\cite{Wang20} introduced an implicit equivariant constraint that enforces CAMs produced by an affine-transformed input to be similar to the affine-transformed CAMs produced by the original input. 

Denote a predefined spatial affine transformation as $T(\cdot)$, the input lobe chunk image as $I$ and $F(I)$ as the dRAM. Equivariant regularization loss $L_{ER}$ can be formulated as
\begin{equation}\label{eq:equivariant_reg}
   \mbox{min:~} ||F(A(I)) - T(F(I)) ||_{1}.
\end{equation}
Equivariant regularization introduces the self-supervising correspondences among affine-transformed images. The affine transformations used in this work combine random resizing ($\left[ 80\%\sim120\%\right]$ of size on each axis) and rotation of 90, 180, or 270 degree at randomly selected axes. 

\subsubsection{dRAM refinement}\label{sec:refine_loss}
Without having the exact lesion percentage per lobe, dRAMs trained by enforcing the predicted percentage to fall into a range may not suffice to delineate lesions accurately. To further improve the segmentation performance, we introduce a refinement step. First, we generate voxel-wise pseudo labels $t^{*}$, where $t^{*}$ is one-hot encoded in a $C \times D \times W \times H$ dimensional matrix. Lesion labels take the overlapping regions between the dRAMs and the candidate lesions (\ref{sec:regio_pro}). Vessel labels are detected vessels from the candidate proposal step. The remaining voxels are considered the background. Because these pseudo labels are generated using automatic methods and may contain errors, we adopt a bootstrapping loss \cite{Reed14} as the refinement loss, which minimizes 
\begin{equation}\label{eq:bootstrapingloss}
\begin{array}{l}
    \sum_{c=0}^{C}[\beta t_{k}^{*} + (1-\beta)z_{c}]log(q_{c}) \\
    z_{c} = 1[c=argmax(q_{i})],i = 0, 1 , \ldots, L,
\end{array}
\end{equation}
where $C$ is the number of classes (3 in our case, including background, vessel, and lesion), $t_{c}$ is the pseudo reference for the class label $c$ and $z_{c}$ is the bootstrapping target produced by the network output. $q_{c}$ is the softmax probability of assigning a voxel into the class $c$. $\beta$ is set to 0.9. The idea of this loss is to leverage the knowledge learned during training to provide hints of the true labels.

The refinement loss and regression loss are trained simultaneously along with the equivariant regularization loss. The total loss is the weighted sum, where the regression loss is weighted 2.0 and others are weighted 1.0. 

\subsubsection{Attention-based dRAM refinement}\label{sec:attention}
To further improve the quality of dRAM, we apply an attention module for refining dRAMs. First, features from the first and second layers of the regression network are squeezed to have $l$ channels via a $1\times 1\times 1$ convolution. Before squeezing, features are detached from the back-propagation, such that refinement training does not update these features. Second, squeezed features are resized to the shape of the input image and concatenated with the input image. 
Denote this concatenated feature map as $x$, $x \in R^{(1+2l)\times D\times W\times H}$. We use $x$ to compute the local voxel-wise affinities in the dRAM $y$, $y \in R^{C \times D\times W\times H}$ ($C$ classes are 0 as background, 1 as vessels, 2 as lesions). The affinity $a(x_{i}, x_{j})$ between two locations $i$ and $j$ in $y$ can be computed via a gated embedded Gaussian function $a(x_{i}, x_{j}) = e^{max((W_{\theta}x_{i})^T(W_{\phi}x_{j}), 0)}$, where $W_{\theta}$ and $W_{\phi}$ are linear transformations to project $x$ into a $l$ dimensional subspace. We eliminate weak affinities using the gate operation by $max(\cdot, 0)$. For each location $i$, only local voxel-wise affinities are measured between $i$ and surrounding locations $j$ in a $3\times 3\times 3$ spatial window taking $i$ as the center within a connectivity of 2, resulting in a total of 18 neighbors. Computing local affinities for all locations in $x$ produces an attention map $A$, $A \in R^{D\times W\times H\times 18}$. For each location, local affinities are normalized within its neighbors $\Omega{j}$. The normalizing factor $\zeta(x)$ can be computed by a summation over the neighboring locations $j$ as $\zeta(x)=\sum_{j\in\Omega{j}}a(x_{i}, x_{j})$. In matrix form, this normalization is equivalent to applying a softmax function over the last dimension on $A$.

To revise the dRAM $y$, we linearly project $y$ into an $l$ dimensional subspace. Then we apply matrix multiplication between the projected dRAM and the attention map $A$ to generate the revised dRAM. This matrix multiplication can be seen as selectively collecting information from its local neighbors for each location in $y$. The impact from local neighbors is determined by the voxel-wise affinities reflecting similarities in terms of low-level image semantics. Finally, we project the revised $y$ back to the input space using a linear transformation $r(\cdot)$. The process of refining the dRAM $y$ at the location $i$ can be formulated as follows:
\begin{equation}\label{eq:att_compute}
    \hat{y_{i}} = r(\frac{1}{\zeta(x)}\sum_{\Omega{j}}a(x_{i}, x_{j})g(y_{j})).
\end{equation}
This attention module, in graph terms, defines a message passing graph where a node (voxel location) connects to its neighbors within a $3 \times 3 \times 3$ grid using connectivity of $2$. And to propagate information among the nodes, we simply matrix multiply the node embeddings (the dRAM $y$) with the attention map ($A$) as \ref{eq:att_compute}.

Considering $A$ as the probability transitional matrix, this attention module acts similarly to the use of random-walks based on local affinity matrix for refining CAMS as used in \cite{Ahn18}. Our method could also have an effect similar to  the deep seeded-region growth method proposed in \cite{Huan18a}, where the authors adopted deep features to define the pairwise similarities between neighboring locations and seeds were defined based on CAMs. The goal of these approaches, in graph terms, is to propagate information among the nodes. Our method enables end-to-end optimization process, whereas random-walks in \cite{Ahn18} and seeded-region growth in \cite{Huan18a} were used as a separate step and not included in the neural network training.

\section{Results}

\subsection{Experimental details}
Training, validation, and testing of each experiment were carried out on a machine with an NVIDIA A100 with 40 GB GPU memory. Train, validation, and test data split can be found in Table \ref{tab:data}. Training each weakly-supervised method took 30 hours, stopped at 200 epochs. Weakly-supervised methods were implemented using Python 3.8, and the Pytorch 1.7.1 library \cite{Pasz19}. Model parameters were initialized according to \cite{He15} and were optimized using stochastic gradient descent with a momentum of 0.9. The initial learning rate was set to $10^{-5}$. For both training and testing of WSS methods, the size of the lobe chunk input images was $80 \times 80 \times 80$ cropped from the input scan resampled to 1.4 millimeters in isotropic spacing. Intensities in input scans were clipped into the range $\left[ -1000\sim400\right]$ HU before re-scaling into $\left[ 0\sim1\right]$. Dense activation maps for separate lobes were tiled together to form a scan-level activation map in the test phase. Then segmentation prediction was obtained by assigning labels with the maximum activation. For the attention module in the proposed method, we set the dimension of the projected subspace to 8 ($l$=8, refer to \ref{sec:attention}). Vessel enhancement filter used for detecting vessels in finding candidate lesions (\ref{sec:regio_pro}) was implemented using the Gaussian scales of 0.8, 1.0, 1.5, 2.0, and 4.0 millimeters, with $\alpha$ equals to 0.5, $\beta$ as 0.5, and $\gamma$ of 15 as correction constants. \par

The fully-supervised experiment was based on the official nn-UNet \cite{Isen20} implementation \footnote{https://github.com/MIC-DKFZ/nnUNet}. We limited the nn-UNet to only operate on 3D full-resolution mode, omitting 2D, cascading, and model ensemble options. We randomly selected 85 scans from the training dataset of 300 scans (Refer to \ref{data_selection}) and let analysts manually labeled lesions as the foreground and the background following the same protocols as annotating the test set. We splitted 85 scans into 65 for training and 20 for validation for training the nn-UNet method. The nn-UNet required training for two full weeks before reaching the required 1000 epochs (Detailed self-configured hyper-parameters listed in the appendix).\par

\subsection{Evaluation Metrics}
The Dice coefficient (DSC), absolute percentage difference (APD), the absolute difference in surface to volume ratio (SVRD), false discovery rate (FDR), and true positive ratio (TPR) were used for measuring the segmentation performance. 

Given the predicted lesion segmentation map $X$ and the segmentation reference $Y$, DSC measures the degree of overlap between the two as:
\begin{equation}
\mbox{DSC}(X, Y) = \frac{2 \times Vol(X\cap Y)}{Vol(X) + Vol(Y)},
\end{equation}
where $\cap$ indicates the intersection between $X$ and $Y$. $Vol(\cdot)$ computes the volume size of the underlying region. 

Using the same notations, APD measures the absolute difference of lesion percentage per lung between the prediction $X$ and the reference $Y$ as:
\begin{equation}
\mbox{APD}(X, Y) = \frac{|Vol(X) - Vol(Y)|}{Vol(Lung)},
\end{equation}
given the volume of the lung as $Vol(Lung)$. APD serves as a volume-based measure. 

The surface-to-volume ratio (SVR) measures the irregularity or compactness of a shape. With the same object size, the ball shape has the smallest SVR, and SVR grows larger when object size increases. SVRD measures the absolute difference in two shapes in terms of the shape irregularity as:
\begin{equation}
\mbox{SVRD}(X, Y) = |\frac{Sur(X)}{Vol(X)} - \frac{Sur(Y)}{Vol(Y)}|,
\end{equation}
where $X$ denotes the predicted lesion mask, $Y$ is the reference mask, and $Sur(\cdot)$ computes the surface area of the underlying region. SVRD serves as a shape-based measurement.

FDR is a precision-based measure that computes the number of falsely-labeled voxels divided by the total number of voxels in prediction as:
\begin{equation}
\mbox{FDR}(X, Y) = 1.0 - \frac{Vol(X\cap Y)}{Vol(X)}.
\end{equation}
between the prediction $X$ and the reference $Y$. 

TPR, or recall, is a measure that indicates the amount of lesions that are detection in the segmentation, computed as:
\begin{equation}
\mbox{TPR}(X, Y) = \frac{Vol(X\cap Y)}{Vol(Y)},
\end{equation}
between the prediction $X$ and the reference $Y$. We compute TPR for each lesion subcategory, namely consolidation, ground-glass, and mixed types.
In addition to segmentation measurements, we also evaluate the classification performance in predicting lobe severity scores.

We translate the predicted lesion percentage per lobe computed as a ratio between segmented lesions and lobes to the lobe severity score using the definition in Table \ref{tab:data}(b). We measure classification accuracy (ACC) to count the number of correctly predicted lobes divided by the total number of lobes in terms of lobe-wise severity. Also, a linearly weighted kappa is computed for measuring the degree of reader agreement using R software package (version 3.6.2; R Foundation for Statistical Computing, Vienna, Austria). 

Besides the quality measurements, the computational complexity was measured by counting the number of Multi-Adds operations (MAC) and the number of network parameters.

\subsection{Comparisons with CAMs and fully-supervised methods}
We denote the regression training with only interval regression loss to obtain segmentation maps as the dRAM method. The proposed method is the extension of dRAM where we trained the regression network with interval regression loss and added equivariant regularization plus refinement loss. 

We compare the proposed method and dRAM method with the CAM-based methods using the same network. The regression network can be turned into a classification network by replacing the regression head with a standard classification head. Given the dense feature maps before the regression head, the classification head computes a mean pooling of dense features per lobe and reshapes the pooled features according to the number of classes using a linear layer. With the classification head, the network can be trained using cross-entropy loss for classifying the presence of lesions. The categorical labels for training the classification network are binary and determined as positive if lobe severity score is non-zero, negative otherwise. CAMs are then generated by multiplying dense feature maps with class-specific weights from the linear layer at the end of the classification. We denote this approach as dCAM because of the use of dense features. The dCAM method is analogous to the MS-CAM method proposed in \cite{Ma20} because U-Net dense features are intrinsically multi-scale, due to the skip-connections between down-sampling and upsampling layer hierarchies. 

Both dCAM and MS-CAM were designed to generate high-resolution CAMs for capturing local details. In order to compare with CAM-based methods where CAMs were generated at a low resolution, we skipped layers in the up-sampling path of the 3D-UNet, transforming the classification network to a network structure typical for generating CAMs where features are convolved in progressively down-sampled resolution. This slimmed classification network is trained for binary classification as that in dCAM, and CAMs are resized by trilinear interpolation to provide lesion segmentation maps at the input resolution. We denote this method as CAM. 
CAMs in the test phase are rescaled to be in the [0,1] range separately for each lobe by eliminating negative values using a Rectified Linear Unit (Relu), and subtracting by the minimal and dividing by the maximum. To generate segmentation maps, rescaled CAMs are binarized using Otsu’s threshold. 

Because raw CAMs produce poor segmentation performance, we designed a simple yet efficient post-processing technique. In post-processing, we apply the lesion proposal technique (described in \ref{sec:regio_pro}) on CAMs by excluding regions outside candidate lesions. The best performing method based on CAMs is the dCAM with post-processing, which is referred to as the baseline method in this paper.
\begin{table*}[!htb]
\caption{Segmentation and classification results using CAM, dCAM, dRAM, and the proposed methods on the test set. nn-UNet indicates the results from the fully-supervised training. Post-processed results are suffixed by '-p'. We use dCAM with post-processing as the baseline method (dCAM-p). Boldface denotes the best result among all WSS methods. The computational complexities in the right column are reported for the worst-case scenario. DSC scores are shown in mean $\pm$ standard deviation.} 
\centering  
\begin{minipage}[t]{\linewidth}
 \centering

\begin{tabular}{c|c|c|c|c|c|c|c|c|c}
\toprule
\multirow{2}{*}{Method} & \multirow{2}{*}{DSC [\%] } & \multirow{2}{*}{APD [\%]} & \multirow{2}{*}{SVRD [\%]} & \multirow{2}{*}{FDR [\%]} & \multicolumn{3}{c}{TPR [\%]} & \multirow{2}{*}{ACC [\%]} & \multirow{2}{*}{MAC/Param} \\
\cline{6-8}
& &  & & & Consolidation & Ground glass & Mixed type&&\\

\midrule
CAM &34.24$\pm$19.59&21.59$\pm$8.79&8.84$\pm$6.14&71.16$\pm$20.37&37.60$\pm$33.00&\textbf{49.79$\pm$27.58}&61.99$\pm$23.37&27.11&\textbf{66.15/7.03}\\\hline
CAM-p&47.29$\pm$20.19&8.03$\pm$9.23&19.27$\pm$11.10&46.09$\pm$24.53&27.71$\pm$25.83&38.41$\pm$24.67&50.51$\pm$20.54&40.89&\textbf{66.15/7.03}\\\hline
dCAM&40.42$\pm$25.56&10.53$\pm$6.77&6.02$\pm$8.14&54.19$\pm$21.66&27.08$\pm$30.75&29.42$\pm$29.86&55.18$\pm$28.99&27.78&462.92/16.32 \\\hline
dCAM-p&48.63$\pm$26.76&7.46$\pm$7.04&15.20$\pm$6.82&34.49$\pm$23.50&28.93$\pm$26.84&35.87$\pm$26.07&52.07$\pm$25.41&42.44&462.92/16.32 \\\hline
dRAM&59.90$\pm$18.75&4.10$\pm$5.26&5.53$\pm$6.21&35.95$\pm$23.92&37.08$\pm$21.37&43.17$\pm$21.23&64.03$\pm$14.65&44.67&462.92/16.32 \\\hline
dRAM-p&58.12$\pm$18.08&6.32$\pm$6.09&9.89$\pm$8.60&28.85$\pm$25.74&32.78$\pm$20.02&36.56$\pm$20.01&55.17$\pm$13.21&41.78&462.92/16.32\\\hline
proposed &\textbf{70.24$\pm$18.66}&\textbf{3.52$\pm$2.78}&\textbf{3.58$\pm$3.19}&\textbf{17.58$\pm$19.15}&\textbf{49.86$\pm$20.80}&48.53$\pm$20.49&\textbf{68.67$\pm$16.81}&\textbf{47.33}&926.41/16.32\\\hline
nnUNet &77.09$\pm$20.67&3.21$\pm$4.77&4.23$\pm$5.75& 23.97$\pm$23.75&71.86$\pm$22.92&75.59$\pm$22.04&84.76$\pm$9.26&50.67&462.92/16.32 \\
\bottomrule
\end{tabular}
\label{tab:camvsdram_qresults}
\end{minipage}
\end{table*}

Quantitative results of CAM, dCAM, dRAM, and the proposed method are shown in Table \ref{tab:camvsdram_qresults}. Without equivariant regularization and refinement loss, training with only interval regression loss (dRAM) reaches 59.90\% in DSC, outperforming all CAM-based methods by a large margin. Using equivariant regularization and refinement loss, the proposed regression training method achieved 70.24\% in DSC, the best weakly-supervised method in comparison. Not surprisingly, a fully-supervised method (nn-UNet) still outperforms weakly-supervised methods, but for several metrics differences with the proposed method are small. Relatively large standard deviations in DSC of all methods are due to the high variation in lesion volume in different cases.  \par

\begin{figure}
    \centering
     \includegraphics[width=0.85\linewidth, height=10cm]{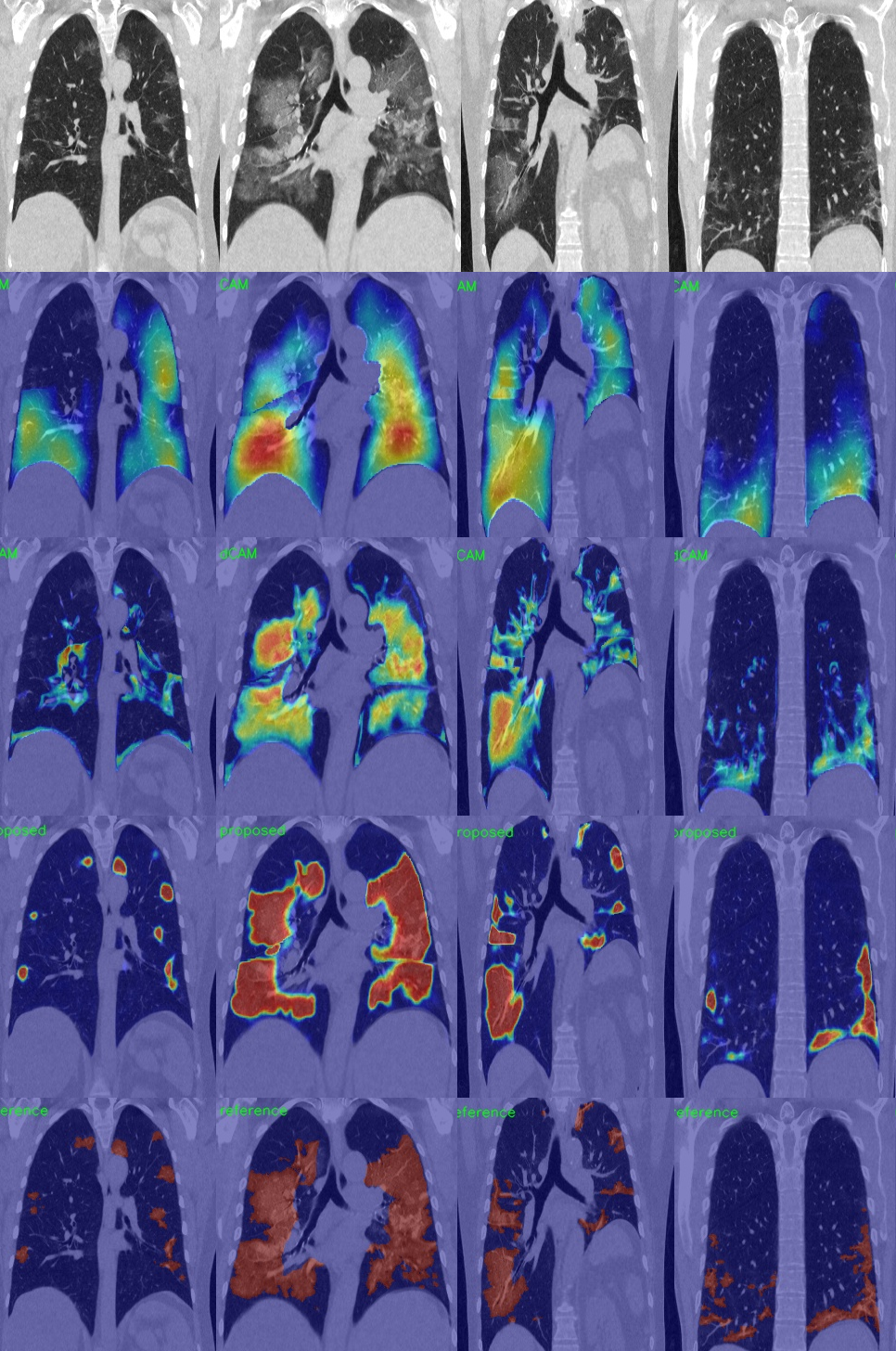}
     \caption{Visualization of CAMs (2nd row), dCAMs (3rd row), and dRAMs in the proposed method (4th row) in one coronal view of four subjects. The 1st and the last row show the original image and the reference standard. We use color map jet for this plot.}
     \label{fig:camvsdcam}
\end{figure}

The post-processing improves the performance of the CAMs-based method, from 40.42\% DSC in dCAM to 48.63\% after post-processing. This is because post-processing reduces a large amount of falsely detected regions in CAMs. In Fig.~\ref{fig:camvsdcam}, CAMs encloses mostly healthy lung parenchyma when there is only a focal lesion (2nd row, 4th column) due to its low-resolution nature. dCAMs, on the other hand, include a large number of false positives in the perivascular area (3rd row). dRAM does not benefit from post-processing, possibly because post-processing may create small disconnected structures by intensity thresholding, which are rarely seen in manual segmentation. This phenomenon is quantitatively viewed as the dramatic increases in SVRD between raw and post-processed methods (visual examples of post-processing shown in the appendix). 

In terms of shape and volume-based measurements, the proposed method reaches 3.52 APD and 3.58 in SVRD, representing a low error rate in predicting lesion volumes and shape compactness compared to manual segmentation. We also notice that the proposed method has a higher voxel-level precision resulting in a lower FDR, but a substantially lower recall (TPRs are lower in all lesion sub-types) than the fully-supervised method. 

In terms of predicting lobe-wise severity scores using the segmentation produced by the listed methods, the proposed method achieved the best classification accuracy at 47.33\% (confusion matrix attached in the appendix), approaching the fully-supervised method at 50.67\%. We also computed the linearly weighted kappa for the baseline, the proposed, and the fully-supervised methods to measure the agreement against a radiologist. Kappa is 46.47(95\%CI:36.02,56.92) for the baseline, 58.61(95\%CI:47.52,69.70) for the proposed, and 56.87(95\%CI:45.99,67.74) for the fully-supervised method, where the proposed method achieves the best kappa. 
% The kappa scores are categorized as slight, fair, moderate, good, or excellent based on k values of 0.20 or less, 0.21–0.40, 0.41–0.60, 0.61–0.80, and 0.81 or higher, respectively, following \cite{Kund03}. There is a moderate agreement between the predicted scores by the proposed method and manual scores, but a fair agreement between the scores predicted by the baseline method and manual scores. 

 Visual inspection of the generated activation maps (Fig.~\ref{fig:camvsdcam}) reveals that blob-like CAMs (2nd row) do not recognize object boundaries as they were was created at low-resolution features. dCAMs (3rd row) produces high-resolution activation maps using dense features, but still suffer from a large amount of over-segmentation and under-segmentation. This can be seen, for example, in the right upper lobe of the first subject (3rd row, 1st column), The activation map in dCAM misses a small pleural lesion, whereas many vessels and opacities near vessels in other subjects are falsely detected. These errors are greatly reduced in the proposed method (4th row). The main reason is that the proposed method was trained to quantify lesions in the image through an interval regression loss. Supplying information about the object size allows the model to learn a much finer feature representation, resulting in a less noisy activation map. This is the major contribution of this work. To our knowledge, this comparison between regression maps and classification maps has not yet been shown in the literature.
 
 In terms of computational complexity, the CAM method based on low-resolution features requires only 66.15GMacs and 7.03M memory for storing model parameters, much less than other methods. The methods based on dense features require 462.92 GMacs and 16.32M GPU memory for storing model parameters. The proposed method needs extra computation for computing the attention map and siamese networks used in the regularization, consuming 926.41 GMacs and 16.32M GPU memory (the memory consumption for the proposed attention module is negligible). In terms of the runtime speed, the proposed method processes each scan at testing time only 2 seconds on average, while the nn-UNet takes roughly 480 seconds to process a scan due to the heavy build-in pre-processing.
 
\subsubsection{Ablation study on dRAM-based methods}

\begin{table}
\caption{Ablation study on the proposed method with or without using the equivariant regularizer (ER), refinement loss (REF), and attention module (AT). The first row represents the performance of the dRAM method, and the final row presents the proposed method. Computational complexities are reported for the worst-case scenario. DSC scores are shown in mean $\pm$ standard deviation.} 
\centering  
\begin{minipage}[t]{\linewidth}
 \centering

\begin{tabular}{c|c|c|c|c|c}
\toprule
ER & REF& AT& DSC,\%& ACC,\%& MAC/Param \\
\midrule
-&-&-&59.90$\pm$18.75&44.67&462.92/16.32 \\\hline
\checkmark&-&-&61.74$\pm$23.62&45.33&925.84/16.32\\\hline
&\checkmark&-&64.62$\pm$22.81&41.33&462.92/16.32\\\hline
\checkmark&\checkmark&-&68.41$\pm$21.59&46.22&925.84/16.32 \\\hline
-&-&\checkmark&62.17$\pm$22.58&45.55&463.49/16.32\\\hline
-&\checkmark&\checkmark&67.72$\pm$21.11&46.00&463.49/16.32\\\hline
\checkmark&\checkmark&\checkmark&\textbf{70.24$\pm$18.66}&\textbf{47.33}&926.41/16.32 \\\hline
\bottomrule
\end{tabular}
\label{tab:abldrams}
\end{minipage}
\end{table}

We conducted an ablation study to evaluate the effectiveness and computational complexity of each component in the proposed WSS framework. These components are the equivariant regularizer (ER), refinement loss (REF), and the proposed attention module (AT). Quantitative results are shown in Table \ref{tab:abldrams}. dRAM trained only with interval regression loss achieves 59.91\% in DSC due to the lack of voxel-level cues. By introducing the dRAM refinement process, we greatly improve DSC to 64.62\% without imposing additional computational burden. The reason is that suppression of vessels and false-detected regions beyond the intensity threshold encourages the network to discover new areas to satisfy the regression target during training. Self-supervised training using the equivariant regularizer contributes to an improvement from 59.91\% DSC to 61.74\% DSC, with a doubled computation cost (925.84 GMacs in the worst-case scenario) during training. The high computation cost is due to the siamese networks taking both original and affine-transformed input images in equivariant regularization. 

\begin{figure}
\centering
\begin{tabular}{ccc}
\includegraphics[width=0.12\textwidth]{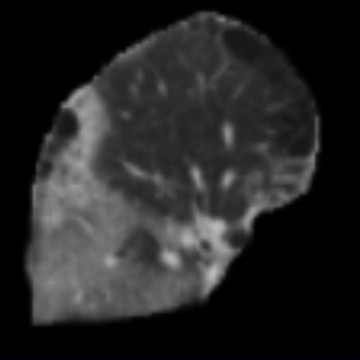}
&\includegraphics[width=0.12\textwidth]{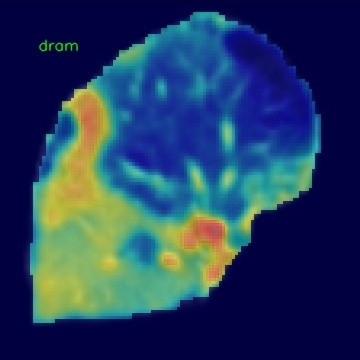}
& \includegraphics[width=0.12\textwidth]{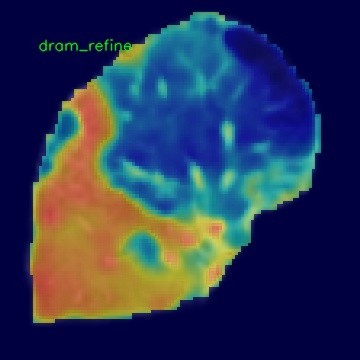}\\
\includegraphics[width=0.12\textwidth]{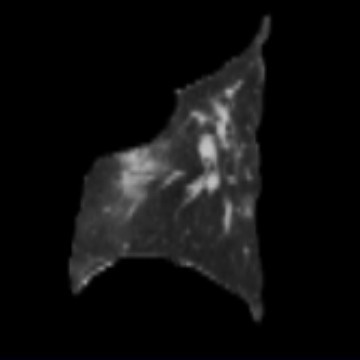}
&\includegraphics[width=0.12\textwidth]{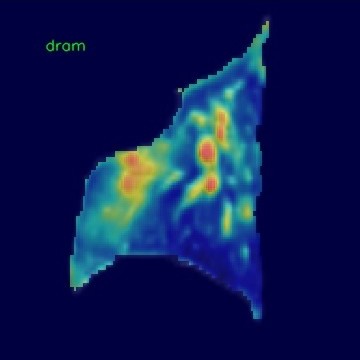}
& \includegraphics[width=0.12\textwidth]{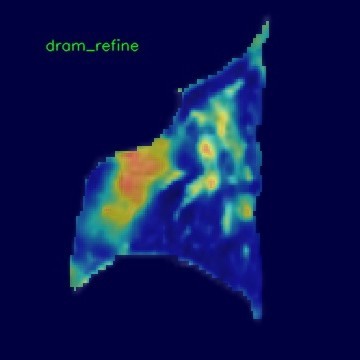}\\
\end{tabular}
\caption{Visualization of dense feature maps before (2nd col) and after (3rd col) the attention module in two input volumes in coronal views (1st col) during training. We use color map jet for this plot.}

\label{fig:at_module}
\end{figure}

The proposed attention module alone boosts the segmentation performance from 59.91\% DSC to 62.17\% DSC. The module updates activation maps at each location selectively using its neighboring information. This is similar to popular CAM-refinement techniques, i.e., conditional random field \cite{Huan18a}, random walkers \cite{Ma20}, and seeded-region growth \cite{Huan18a}. Fig.~\ref{fig:at_module} shows the activation maps before and after the attention module in two input images during training. We notice that neighboring information is gathered and weighted more towards high attenuation areas. This makes low attenuation regions (probably healthy lung parenchyma) less involved in further computation, suppressing false alarms and forcing the network to look for other regions for finding lesions.
With the equivariant regularizer, refinement loss, and the attention module, the proposed method ultimately reaches a DSC of 70.24\%. Even though the computation cost of the proposed method is high, we argue that the large computational burden only occurs during training. For testing, all listed methods in the ablation study are the same. 

All listed approaches do not differ much in terms of classification performance, where the proposed method achieves the best accuracy at 47.33\%. 

\begin{figure}
\centering
\includegraphics[width=0.85\linewidth]{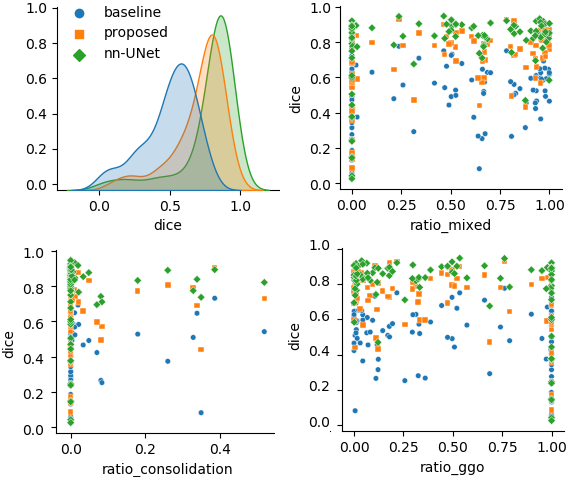}
\caption{Robustness study on different lesion subtypes for the baseline, the proposed, and fully-supervised methods. The top-left image shows the overall DSC distribution on the test set. The rest images show the correlation between the lesion percentage of a specific subtype and corresponding DSC. }
\label{fig:robustness}
\end{figure}
\begin{figure*}[ht!]
\centering
\setlength{\tabcolsep}{0.001\textwidth}
\includegraphics[width=0.9\linewidth, height=15cm]{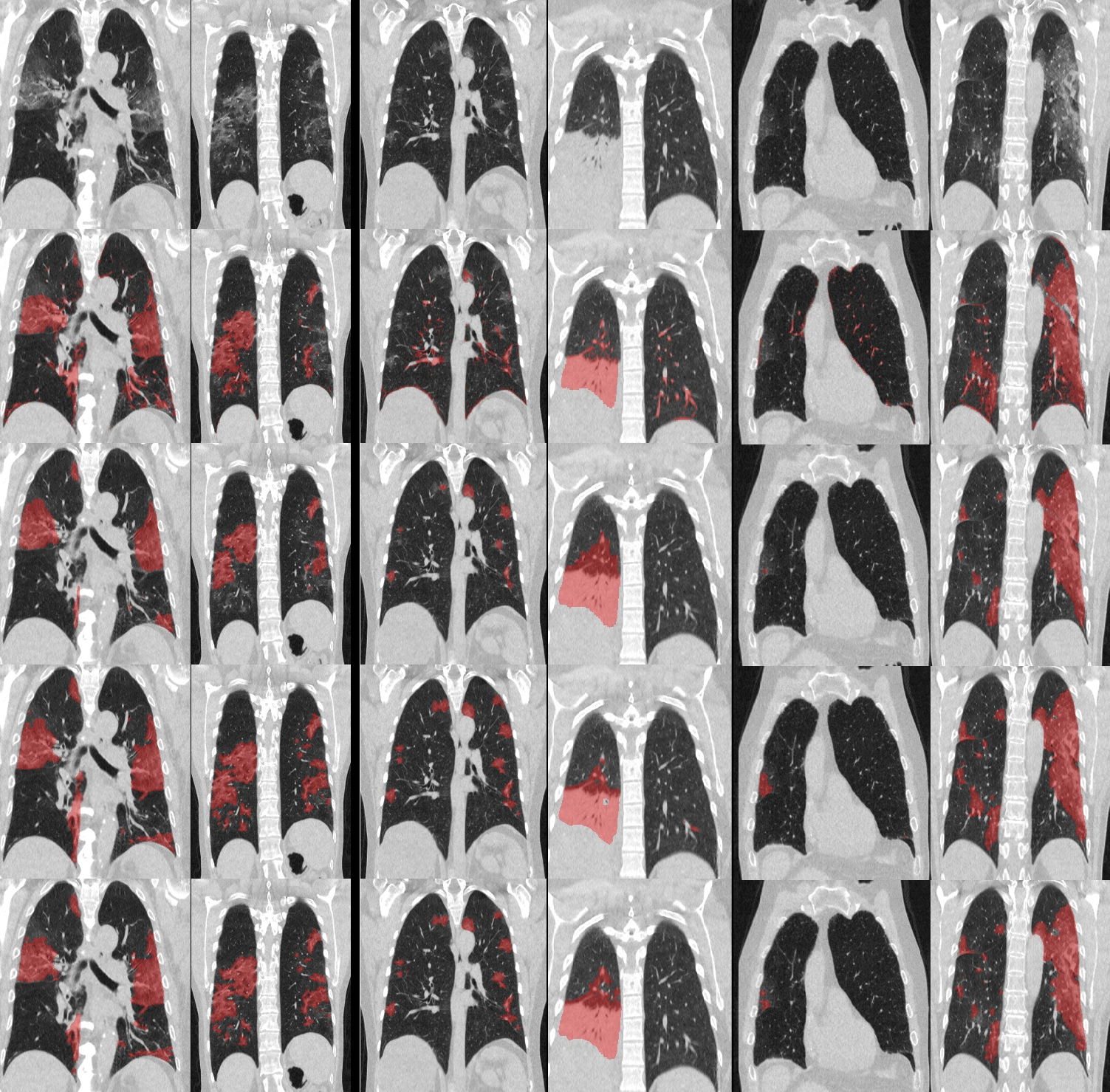}
\caption{Segmentation results for six representative test cases in coronal views (1st row) represented in columns. The 2nd, 3rd, 4th, and 5th rows show the segmentation results of the baseline, the proposed, nnU-Net method, and the segmentation reference respectively. } \label{fig:qresult}
\end{figure*}
\subsection{Robustness study on lesion subtypes}
This study aims to examine if the segmentation method performs worse or better for a specific lesion subtype. We computed the correlation between the volume ratio of each subtype and the segmentation performance (in DSC,\%) on every test image. Each point in Fig.~\ref{fig:robustness} tallies the correlation on each test image, where the top-left image shows the distribution of DSC scores for the baseline, the proposed, and the fully-supervised methods. The other images rank the percentage of each lesion subtype (the volume of each lesion subtype divided by the overall lesion volume) at $x$-axis, and $y$-axis shows the DSC scores.

The top-right subplot shows that all methods perform well on mixed lesions because points are more concentrated at the top right corner. However, the performance is relatively weak for the ground-glass and consolidated lesions because points are densely distributed at the top-left corner on the subplots corresponding to the ground-glass and consolidation subtypes.

\subsection{Qualitative Results}
As shown in Fig.~\ref{fig:qresult}, the result from the baseline method (2nd row) exhibit substantial under-segmentation in pleural lesions (2nd, 3rd, 5th, 6th columns). Also, the baseline method often over-segments perivascular regions (3rd, 4th columns). The proposed method (3rd row) generally performed well on lesion segmentation. However, compared with the nn-UNet results, the proposed method under-segments small pleural ground-glass lesions (3rd and 5th column). One reason is that the lobe-wise severity scores only represent an interval of the per-lobe lesion percentage and not the exact percentage, which potentially allows the network to tolerate certain mistakes. On the other hand, ground-glass opacities may create challenges in visual recognition. This challenge may also cause measurement errors for radiologists in labeling severity scores, further contributing to confusion regarding ground-glass opacities in our methods.

% \section{Discussion}
% \input{sections/discussion}

\section{Discussion and Conclusion}
We proposed a novel weakly-supervised segmentation method. This method can train a segmentation network using only severity scores provided for individual lobes, where these scores correspond to a range of percentages of affected regions in these lobes. From only these lobe scores, the network can generate dense regression activation maps (dRAMs). These dRAMs were refined by the proposed attention module using low-level semantics. The proposed attention module enriches the semantic representation at each voxel based on its local neighbors (affinities). 

The proposed method achieved significant improvements in segmentation performance compared with the baseline approach. In terms of the prediction of lobe severity scores, the proposed method reached a moderate agreement with the scores assigned by the radiologist, while the baseline method only reached a fair agreement. Our results showed that the proposed model sometimes under-segments small ground-glass lesions in the subpleural region of the lung periphery. However, as weak labels are cheap to collect, more advanced approaches can be built upon our model using our methods as the initial seed for interactive (e.g., adaptive learning scenarios) or iterative refinement (e.g., knowledge distillation). 

The proposed method is generic and can be easily adapted to other weakly-supervised segmentation problems if specific object statistics are given and can be used as the regression target. Visually assessing the amount of affected regions is common in radiology, and therefore we believe that the proposed weakly-supervised segmentation framework can be used for many segmentation problems in medical imaging, where automatic segmentation is often used for quantification analysis. In these scenarios, our framework can directly use visually assessed quantification results from radiological scoring systems as the regression target.

\bibliographystyle{plain}
\bibliography{fullstrings,medlinestrings,main}

\begin{thebibliography}{10}

\bibitem{Ahn18}
Jiwoon Ahn and Suha Kwak.
\newblock Learning pixel-level semantic affinity with image-level supervision
  for weakly supervised semantic segmentation.
\newblock pages 4981--4990, 2018.

\bibitem{Cice16}
{\"O}.~{\c{C}}i{\c{c}}ek, A.~Abdulkadir, S.~S. Lienkamp, T.~Brox, and
  O.~Ronneberger.
\newblock 3{D} {U}-{N}et: Learning dense volumetric segmentation from sparse
  annotation.
\newblock pages 424--432, 2016.

\bibitem{Fan20}
Deng~Ping Fan et~al.
\newblock {Inf-Net}: Automatic {COVID}-19 lung infection segmentation from {CT}
  images.
\newblock 39:2626--2637, 2020.

\bibitem{Fran98}
A.~F. Frangi, W.~J. Niessen, K.~L. Vincken, and M.~A. Viergever.
\newblock Multiscale vessel enhancement filtering.
\newblock volume 1496, pages 130--137, 1998.

\bibitem{Gonz20}
Cristina Gonz\'{a}lez-Gonzalo, Bart Liefers, Bram van Ginneken, and Clara~I.
  S\'{a}nchez.
\newblock Iterative augmentation of visual evidence for weakly-supervised
  lesion localization in deep interpretability frameworks: application to color
  fundus images.
\newblock 39(11):3499--3511, 2020.

\bibitem{He15}
K.~He, X.~Zhang, S.~Ren, and J.~Sun.
\newblock Delving deep into rectifiers: Surpassing human-level performance on
  imagenet classification.
\newblock pages 1026--1034, 2015.

\bibitem{Huan18a}
Zilong Huang, Xinggang Wang, Jiasi Wang, Wenyu Liu, and Jingdong Wang.
\newblock Weakly-supervised semantic segmentation network with deep seeded
  region growing.
\newblock pages 7014--7023, 2018.

\bibitem{Isen20}
Fabian Isensee, Paul~F. Jaeger, Simon A.~A. Kohl, Jens Petersen, and Klaus~H.
  Maier-Hein.
\newblock nn{U}-{N}et: a self-configuring method for deep learning-based
  biomedical image segmentation.
\newblock {\em Nature methods}, 18(2):203--211, 2020.

\bibitem{Ji19}
Zhanghexuan Ji, Yan Shen, Chunwei Ma, and Mingchen Gao.
\newblock Scribble-based hierarchical weakly supervised learning for brain
  tumor segmentation.
\newblock volume 11766, pages 175--183, 2019.

\bibitem{Lara20}
Issam Laradji et~al.
\newblock A weakly supervised region-based active learning method for
  {COVID}-19 segmentation in {CT} images.
\newblock {\em arXiv:2007.07012}, 2020.

\bibitem{Less20}
Nikolas Lessmann et~al.
\newblock Automated assessment of {COVID}-19 reporting and data system and
  chest {CT} severity scores in patients suspected of having {COVID}-19 using
  artificial intelligence.
\newblock 298(1):E18--E28, 2021.

\bibitem{Li20}
Kunhua Li et~al.
\newblock The clinical and chest {CT} features associated with severe and
  critical {COVID}-19 pneumonia.
\newblock 55(6):327--331, 2020.

\bibitem{Ma20}
Xiao Ma, Zexuan Ji, Sijie Niu, Theodore Leng, Daniel~L. Rubin, and Qiang Chen.
\newblock {MS}-{CAM}: Multi-scale class activation maps for weakly-supervised
  segmentation of geographic atrophy lesions in {SD-OCT} images.
\newblock 24:3443--3455, 2020.

\bibitem{Otsu79}
N.~Otsu.
\newblock A threshold selection method from gray level histograms.
\newblock 9(1):62--66, 1979.

\bibitem{Papa15}
George Papandreou, Liang-Chieh Chen, Kevin~P. Murphy, and Alan~L. Yuille.
\newblock Weakly-and semi-supervised learning of a deep convolutional network
  for semantic image segmentation.
\newblock pages 1742--1750, 2015.

\bibitem{Pasz19}
Adam Paszke et~al.
\newblock Pytorch: An imperative style, high-performance deep learning library.
\newblock pages 8024--8035, 2019.

\bibitem{Pinh15}
Pedro~O Pinheiro and Ronan Collobert.
\newblock From image-level to pixel-level labeling with convolutional networks.
\newblock pages 1713--1721, 2015.

\bibitem{Prok20}
Mathias Prokop et~al.
\newblock {CO-RADS} - a categorical {CT} assessment scheme for patients with
  suspected {COVID-19}: definition and evaluation.
\newblock 296(2):E97--E104, 2020.

\bibitem{Reed14}
Scott Reed, Honglak Lee, Dragomir Anguelov, Christian Szegedy, Dumitru Erhan,
  and Andrew Rabinovich.
\newblock Training deep neural networks on noisy labels with bootstrapping.
\newblock {\em arXiv:1412.6596}, 2014.

\bibitem{Roth19}
Holger Roth et~al.
\newblock Weakly supervised segmentation from extreme points.
\newblock volume 11851, pages 42--50, 2019.

\bibitem{Shi20a}
Heshui Shi et~al.
\newblock Radiological findings from 81 patients with {COVID}-19 pneumonia in
  {Wuhan}, {China}: a descriptive study.
\newblock 20(4):425--434, 2020.

\bibitem{Wang20a}
Xinggang Wang et~al.
\newblock A weakly-supervised framework for {COVID}-19 classification and
  lesion localization from chest {CT}.
\newblock 39(8):2615--2625, 2020.

\bibitem{Wang20}
Yude Wang, Jie Zhang, Meina Kan, Shiguang Shan, and Xilin Chen.
\newblock Self-supervised equivariant attention mechanism for weakly supervised
  semantic segmentation.
\newblock pages 12275--12284, 2020.

\bibitem{Wei17a}
Yunchao Wei et~al.
\newblock {STC}: A simple to complex framework for weakly-supervised semantic
  segmentation.
\newblock 39(11):2314--2320, 2017.

\bibitem{Wei17}
Yunchao Wei, Jiashi Feng, Xiaodan Liang, Ming-Ming Cheng, Yao Zhao, and
  Shuicheng Yan.
\newblock Object region mining with adversarial erasing: A simple
  classification to semantic segmentation approach.
\newblock pages 1568--1576, 2017.

\bibitem{Xie20}
Weiyi Xie, Colin Jacobs, Jean-Paul Charbonnier, and Bram van Ginneken.
\newblock Relational modeling for robust and efficient pulmonary lobe
  segmentation in {CT} scans.
\newblock 39:2664--2675, 2020.

\bibitem{Xu20a}
Zhanwei Xu et~al.
\newblock {GASNet}: Weakly-supervised framework for {COVID}-19 lesion
  segmentation.
\newblock {\em arXiv:2010.09456}, 2020.

\bibitem{Yao20}
Qingsong Yao, Li~Xiao, Peihang Liu, and S~Kevin Zhou.
\newblock Label-free segmentation of {COVID}-19 lesions in lung {CT}.
\newblock {\em arXiv:2009.06456}, 2020.

\end{thebibliography}

\newpage
\newpage
\appendix

\subsection{nn-UNet settings}
\begin{table}[hbt!]
\caption{Main settings chosen by the nnUNet framework to train the segmentation network with manual annotations (fully-supervised training). See \cite{Isen20} for more information.} 
    \centering
    \begin{tabular}{c|c}
    parameters & values \\
    \hline
         number of pooling operations in z,x,y direction & [5, 5, 4]\\
         number of features after first conv & 32\\
         conv per stage &  2\\
         optimizer & SGD (momentum: 0.99) \\
         initial learning rate & 0.01 \\
         number of epochs & 1000  \\
         number of batches used in every epoch  & 250 \\
         number of images per batch & 2 \\
         patch size in z,y,z direction  & [128 160 112] \\
         normalization schemes & (0, 'CT')
    \end{tabular}
    
    \label{tab:nnunet}
\end{table}

\subsection{Effect of post-processing}
\begin{figure}[hbt!]
    \centering
     \includegraphics[width=0.85\linewidth]{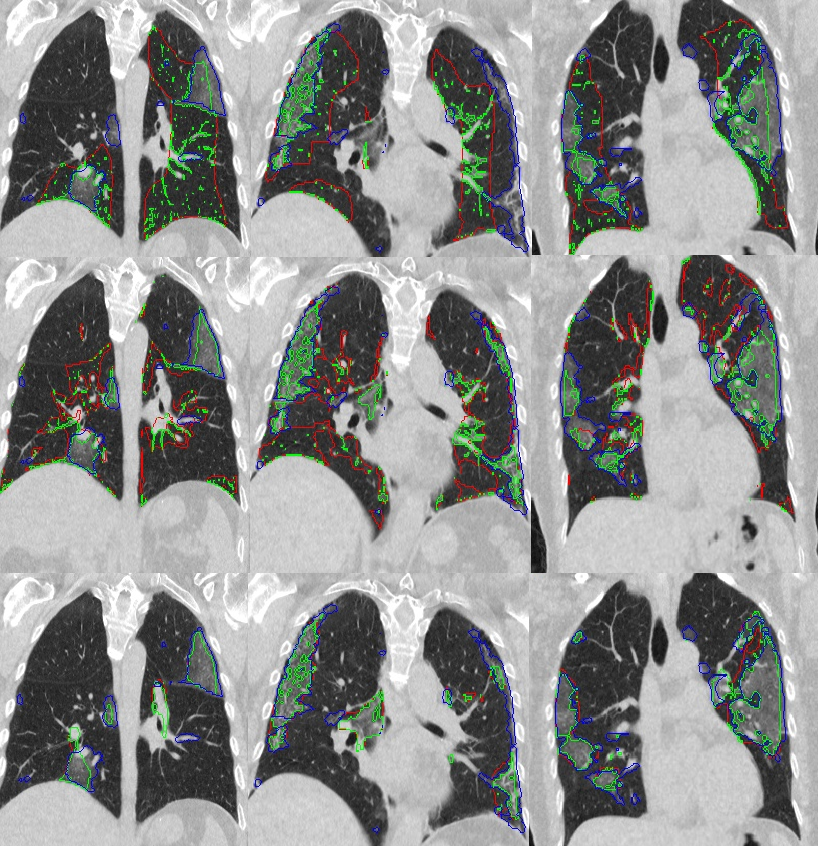}
     \caption{Post-processed segmentation results of CAM (1st row), dCAM (2nd row), and dRAM (3rd row) methods in three different test subjects listed column-wise. Segmented contours from raw activation maps were drawn in \textcolor{CRED}{\rule{.2cm}{.2cm}}, post-processed activation maps in \textcolor{CGREEN}{\rule{.2cm}{.2cm}}, and reference masks in \textcolor{CBLUE}{\rule{.2cm}{.2cm}}. Contours were drawn in orders of raw, post-processed, and the reference. Best viewed in color.}
     \label{fig:post_effect}
\end{figure}

\subsection{Confusion Matrix of the proposed method in predicting lobe-wise severity scores}
\begin{figure}[hbt!]
    \centering
     \includegraphics[width=0.85\linewidth]{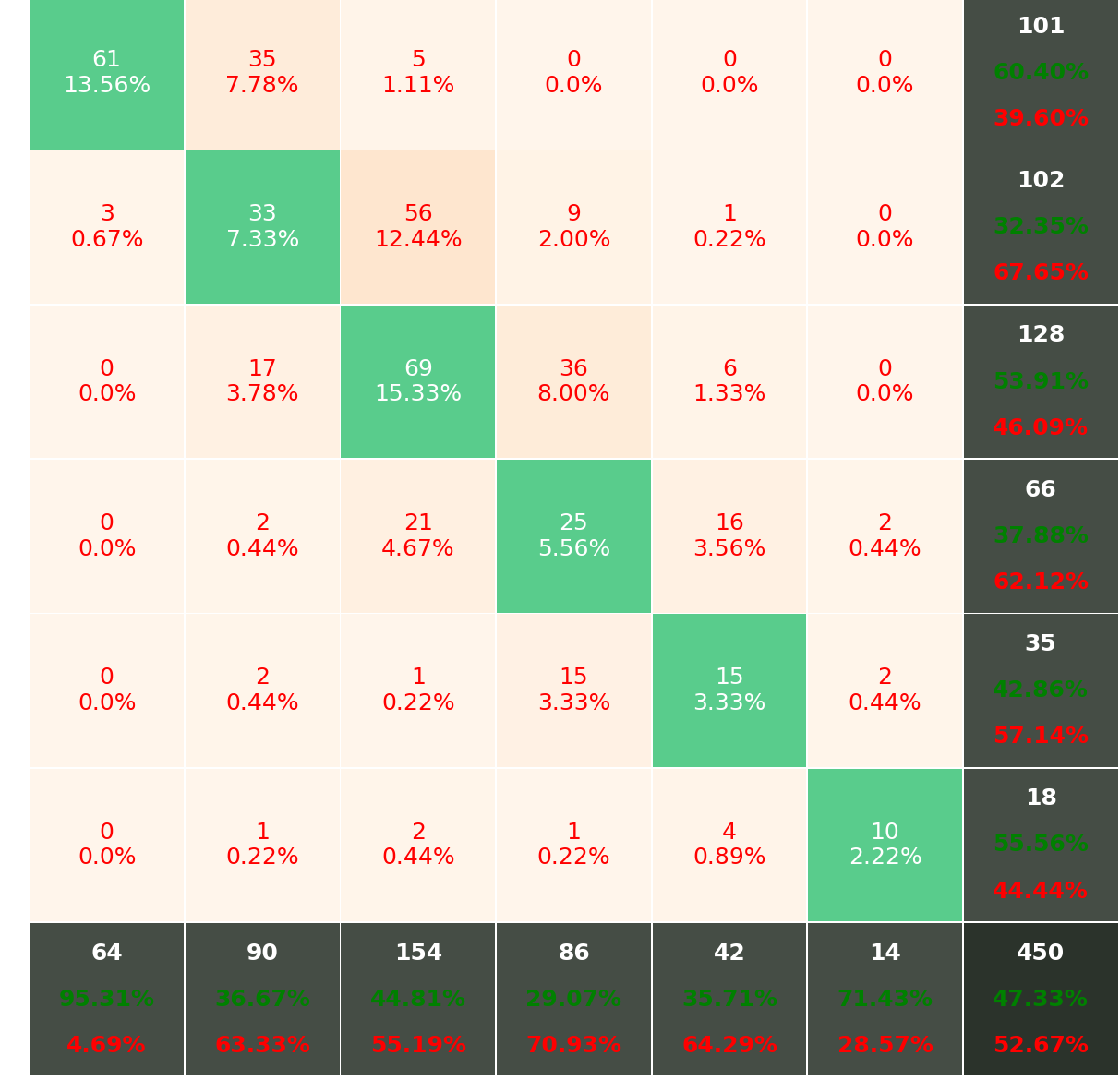}
     \caption{Confusion matrix of the proposed method in predicting lobe-wise severity scores on the test set, rows represent predicted score and columns show the target scores.}
     \label{fig:cm}
\end{figure}
\end{document}